\documentclass[
 reprint,
 amsmath,amssymb,
 aip,apl,
]{revtex4-1}

\usepackage{graphicx}
\usepackage{bm}
\usepackage{hyperref}
\usepackage{placeins}
\usepackage{siunitx}
\usepackage{longtable}
\usepackage{multirow}
\usepackage{booktabs}
\usepackage{dcolumn}
\usepackage[normalem]{ulem}

\hypersetup{breaklinks=true} 
\usepackage{graphicx}
\usepackage{xcolor}

\begin{document}

\preprint{APS/123-QED}

\title{Low-loss Material for Infrared Protection of Cryogenic Quantum Applications}

\author{Markus Griedel}
\author{Max Kristen}
\affiliation{Physikalisches Institut, Karlsruhe Institute of Technology, 76131 Karlsruhe, Germany}
\affiliation{Institute for Quantum Materials and Technologies, Karlsruhe Institute of Technology, 76021 Karlsruhe, Germany}

\author{Biliana Gasharova}
\author{Yves-Laurent Mathis}
\affiliation{Institute for Beam Physics and Technology, Karlsruhe Institute of Technology, 76021 Karlsruhe, Germany}

\author{Alexey V. Ustinov}
\author{Hannes Rotzinger}
\email{rotzinger@kit.edu}
\affiliation{Physikalisches Institut, Karlsruhe Institute of Technology, 76131 Karlsruhe, Germany}
\affiliation{Institute for Quantum Materials and Technologies, Karlsruhe Institute of Technology, 76021 Karlsruhe, Germany}

\date{\today}

\begin{abstract}

The fragile quantum states of low-temperature quantum applications require protection from infrared radiation caused by higher-temperature stages or other sources. {In particular, signal lines have to be manufactured to prevent infrared photons entering through dielectric openings while maintaining low microwave loss.}
We propose a material system that can efficiently block radiation up to the optical range while transmitting photons at low gigahertz frequencies. It is based on the effect that incident photons are strongly scattered when their wavelength is comparable to the size of particles embedded in a weakly absorbing medium (Mie-scattering). \\
The goal of this work is to tailor the absorption and transmission spectrum of an non-magnetic epoxy resin containing sapphire spheres by simulating its dependence on the size distribution.  Additionally, we fabricate several material compositions, characterize them, as well as other materials, at optical, infrared, and gigahertz frequencies. 
In the infrared region (stop band) the attenuation of the Mie-scattering optimized material is high and comparable to that of other commonly used filter materials.  At gigahertz frequencies (pass-band), the prototype filter exhibits a high transmission at millikelvin temperatures, with an insertion loss of less than $0.4$\,dB below 10\,GHz. 

\end{abstract}

\maketitle
\begin{figure}
        \includegraphics[width=0.48\textwidth]
        {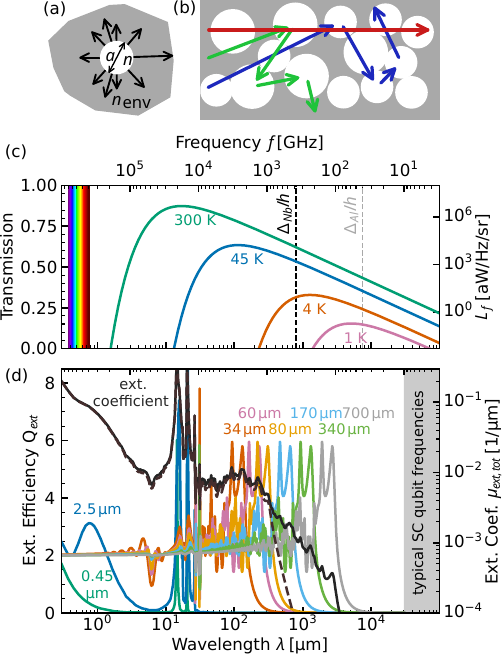}
    \caption{\label{fig:introduction}(a) Scattering at a single sphere ($n=3.69$; $x=1.5$) in an epoxy resin matrix ($n_{env}=1.5$). (b) Spheres of different sizes in a epoxy matrix with scattering path for different wavelength from short (blue) to long (red) wavelength. (c) Solid colored lines represent Planck radiation spectrum of the different cryogenic stages (right axis, $L_f$); dashed lines indicate the superconducting gap frequencies for Niobium (black) and Aluminum (gray). The visible light spectrum is indicated on the far left. (d) Calculated Mie scattering for sapphire spheres of varying sizes in an epoxy resin matrix as a function of wavelength (left axis). The black solid (dashed) line is the total extinction efficiency of the composition SP0.45-700 (SP0.45-80) of different sphere sizes (right axis).}
\end{figure}

The low noise environment at ultra low temperatures in the millikelvin range is a key requirement for many quantum applications, such as semiconducting or superconducting qubits. In particular, the low dissipation of superconducting devices stems from the superconducting energy gap, which suppresses low energy excitations. However, when incident photons with larger energy are absorbed, Cooper pairs get broken, which introduces losses and, consequently, noise. Depending on the application, the performance of superconducting circuits can then degrade severely. For instance, SQUID sensors show a lower magnetic flux sensitivity, or qubits a reduced coherence and energy relaxation time\cite{wellstoodLowfrequencyNoiseDc1987,Drung_2011,barendsMinimizingQuasiparticleGeneration2011,devisserMicrowaveinducedExcessQuasiparticles2012,
wangMeasurementControlQuasiparticle2014,
gustavssonSuppressingRelaxationSuperconducting2016,Mannila_2022,liuQuasiparticlePoisoningSuperconducting2024,
benevidesQuasiparticleDynamicsSuperconducting2024}. The spectrum of unwanted radiation ranges from optical to infrared (IR) wavelength stemming from room temperature sources, but also from lower temperature stages of the cryostat; see Fig.~\ref{fig:introduction}(c) for the Planck spectra of several temperature stages. To prevent excess photons, devices must be carefully shielded and are usually enclosed in metallic light-tight boxes, often with several layers\cite{barendsMinimizingQuasiparticleGeneration2011, Mannila_2022}.

Electrical leads to the superconducting device require a dielectric insulation from the shield. For operation frequencies up to a few tens of GHz, coaxial cables are commonly used, with the inner wire insulated from the outer conductor by a dielectric material such as
polytetrafluoroethylene (PTFE) or high density polyethylene (HDPE). A drawback of the insulator is that it creates an opening in the otherwise light-tight shield, allowing IR photons to enter the box. An ideal solution to this problem would be to add a low-pass filter to the coaxial cable that ensures lossless transmission at operating frequencies and efficient blocking of IR photons. However, the large range of unwanted photons, which can span over five orders of magnitude up to the far infrared (FIR) range, makes this a challenging task. Widely used approaches\cite{lukashenkoImprovedPowderFilters2008a,
kreikebaumOptimizationInfraredMagnetic2016,danilinEngineeringMicrowaveInfrared2022,
paquetteAbsorptiveFiltersQuantum2022,ivanovRobustCryogenicMatched2023} employ light-absorbing materials such as Eccosorb~\cite{lairdEccosorbrCRSDatasheet2015}.
These materials are optimized for efficient, frequency-independent absorption of radiation. While this solution can be highly effective in the blocking band, it also attenuates the signal in the desired pass band.

In this paper, we describe an approach in which the properties of a compound of high-quality dielectric sapphire spheres embedded in a polymer (epoxy resin) matrix. This solution is tailored to achieve both goals, a very high absorption of infrared radiation and a very low absorption at gigahertz frequencies. The spheres have varying diameters ranging from hundreds of nanometers to hundreds of micrometers. The key idea is that incident radiation interacts strongly with the spheres when the wavelength is comparable to the sphere's diameter, i.e., employing the phenomenon known as Mie scattering. At long wavelengths (several millimeters to centimeters), the interaction is minimal and the wave can pass through{\cite{essinger-hilemanAerogelScatteringFilters2020,helsonNovelInfraredblockingAerogel2022}.} We favor a dielectric material system, that is non-magnetic and does not contain any electrically conductive constituents to keep the overall loss low. In particular, sapphire is known for its low microwave loss~\cite{krupkaUseWhisperinggalleryModes1999, kudraHighQualityThreedimensional2020}.

The paper is structured as follows. First, the basic principle is introduced, along with simulations based on the Mie theory for strong light–matter interaction. The second part presents and discusses measurement data on the absorption of IR by sapphire spheres embedded in a dielectric matrix. These data are then compared with that of other commonly used materials, such as PTFE, HDPE (transparent and black), Eccosorb CR 124, Stycast 2850 FT, UHU plus Endfest 300 (the primarily used epoxy resin) and UHU plus Endfest 300 loaded with stainless steel or copper powder. Supplementary materials contain further details. 
Finally, we focus on GHz frequency transmission at cryogenic temperatures and present the characteristics of our prototype IR filter.

The classical Beer–Lambert law\cite{Beer_1852} $I/I_0 = \exp(-\mu l)$ describes the extinction of radiation with the intensity of outgoing radiation relative to the incoming radiation in media where the interaction is relatively weak and the wavelength of the radiation differs from the internal structure. Here, $l$ is the length of the medium. The extinction coefficient $\mu$ can be approximated by $\mu_{abs} + \mu_{sca}$ when considering the absorption and scattering of incident radiation. Due to scattering, the radiation is dispersed in the material (Rayleigh scattering), with a fraction being backscattered. Absorptive losses are converted to heat. 

Strong scattering of the incident radiation can be observed when the size parameter $x =\pi a/\lambda$ of the scatterer of diameter $a$ in the medium is on the order of unity. The physics of these processes was worked out by G. Mie and P. Debye and can be found in numerous textbooks (e.g. Ref.~\onlinecite{Kerker_1969,bohrenAbsorptionScatteringLight1998}). 

The total extinction cross section for the particle comprises the energy abstracted from the incident beam by both scattering and absorption $C_\mathrm{ext} = C_\mathrm{sca} + C_\mathrm{abs}$ \cite{Kerker_1969} and can be calculated numerically. Our simulation results discussed below are based on the powerful MiePython\cite{prahlMiepythonPurePython2024} software package. 
Figure~\ref{fig:introduction}(a) illustrates a calculation of the angle distribution of the scattered radiation. A sapphire sphere ($n = 3.69, a = 50$~µm), embedded in a dielectric matrix (epoxy\cite{uhuEndfest300Datenblatt2025}, $n \approx 1.5$), reacts to incident radiation at $\lambda = 100$~µm by scattering photons almost isotropically. Figure~\ref{fig:introduction}(d) shows the extinction efficiency, $Q_\mathrm{ext} = C_\mathrm{ext}/\pi (a/2)^2$, for several sphere diameters ranging from 0.45\,$\mu$m to 700\,$\mu$m (individual colors), as a function of the incident wavelength. Except for very small spheres, the general behavior is similar. At wavelengths below 10\,$\mu$m, $Q_\mathrm{ext}$ is approximately independent of wavelength. The vibrational mode of sapphire at 10–20\,$\mu$m are common to all sphere diameters\cite{querryOpticalConstants1985}, whereas the position of the resonance at longer wavelengths depends on the individual $a$ and thus on $x$. Please note the steep decay\cite{bohrenAbsorptionScatteringLight1998} of the $Q_\mathrm{ext}$ curves at {$x\ll1$} which drops of as $\lambda^4$ (Rayleigh limit).

We now make the following approximations\cite{Kerker_1969,bohrenAbsorptionScatteringLight1998}, which are illustrated in Fig.~\ref{fig:introduction}(b): If radiation is scattered by several spheres with the same $a$, the total $Q_\mathrm{ext}$ should be additive.
Similarly, we assume that the total $Q_\mathrm{ext}$ for a medium with a distribution of $a$ will increase the extinction bandwidth\cite{bohrenAbsorptionScatteringLight1998}. This is shown by the solid black line in Fig.~\ref{fig:introduction}(d),  which expresses the expected extinction coefficient $\mu_\mathrm{ext,tot}$ in units of $\si{1/\micro\meter}$, given by 
\begin{align}
    \mu_\mathrm{ext,tot} &= N_\mathrm{tot} C_\mathrm{ext,tot} 
    \label{eq:extinction_tot}
\end{align}
where $N_\mathrm{tot}$ is the total number density per volume {given for a mixture of equal masses by $N_\mathrm{tot}=\sum_i 1/(\xi V_i (1+ \rho_\mathrm{sapphire}/\rho_\mathrm{epoxy}))$}, and $C_\mathrm{ext,tot} = \sum_i \gamma_i Q_{\mathrm{ext},i} \pi (a/2)^2$ the total extinction cross section; $\xi$ is the number of different spheres in the mixture; the density ratio $\rho_\mathrm{sapphire}/\rho_\mathrm{epoxy} \approx 3.6$. The calculation of $C_\mathrm{ext}$ also includes a weighting factor $\gamma_i=V_i/\sum_j V_j$ which takes into account the relative number of individual spheres per volume.

We use commercially available\cite{finaladvancedmaterials2MS001FINALAdvanced2024} sapphire powders (SP), Table~\ref{tab:parameters} lists the median particle diameters assuming a spherical shape, the corresponding wavelength $\lambda_{x'}$ where  $x=1$, the real and imaginary part of the refractive index\cite{querryOpticalConstants1985} $n(\lambda)$, $\kappa(\lambda)$, and weighting factor $\gamma$. The primary mixtures investigated are SP0.45-80 ($\xi=5$) and SP0.45-700 ($\xi=8$). All investigated sapphire-epoxy composites are non-transparent in the optical range.

\begin{table}[h]
    \centering
    \caption{\label{tab:parameters}Parameters for sapphire spheres that are investigated. The wavelength $\lambda_{x'}$, refractive index components $n(\lambda_{x'})$ and $\kappa(\lambda_{x'})$ are given for $x =1$. $\gamma$ is specific for the SP0.45-700 mixture.}
    \begin{ruledtabular}
    \begin{tabular*}{0.48\textwidth}{@{\extracolsep{\fill}} r r c S S }
        {$a$ [$\mu\mathrm{m}$]} & {$\lambda_{x'}$ [$\mu\mathrm{m}$]} & {$n(\lambda_{x'})$} & {$\kappa(\lambda_{x'})$} & {$\gamma$} \\
        \hline
        0.45 & 1.41   & 1.73 & {\(1.80 \times 10^{-2}\)} & {\(9.9 \times 10^{-1}\)} \\
        2.5  & 7.85   & 1.36 & {\(3.43 \times 10^{-2}\)} & {\(5.8 \times 10^{-3}\)} \\
        34   & 106.8 & 3.69 & {\(3.00 \times 10^{-2}\)} & {\(2.3 \times 10^{-6}\)} \\
        60   & 188.5 & 3.69 & {\(3.00 \times 10^{-2}\)} & {\(4.2 \times 10^{-7}\)} \\
        80   & 251.3 & 3.69 & {\(3.00 \times 10^{-2}\)} & {\(1.8 \times 10^{-7}\)} \\
        170  & 534.1 & 3.69 & {\(3.00 \times 10^{-2}\)} & {\(1.8 \times 10^{-8}\)} \\
        340  & 1068 & 3.69 & {\(3.00 \times 10^{-2}\)} & {\(2.3 \times 10^{-9}\)} \\
        700  & 2199 & 3.69 & {\(3.00 \times 10^{-2}\)} & {\(2.6 \times 10^{-10}\)} \\
    \end{tabular*}
    \end{ruledtabular}
\end{table}

\begin{figure}
    \centering
    \includegraphics[width=0.48\textwidth]
            {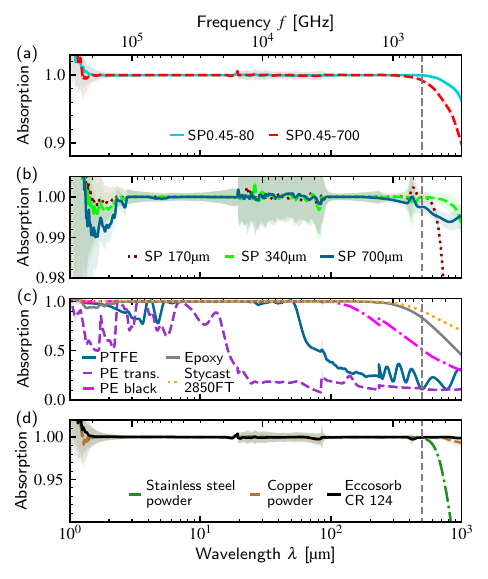}
    \caption{\label{fig:IR_absorption}
    Absorption spectra of various materials at a thickness of \SI{1.5}{\milli\meter}. Please note the different scales. {The gray dashed line illustrates the region shown in Fig.~\ref{fig:extinction_combined}}.
    (a) Sapphire powder mixtures SP0.45-80 and SP0.45-700, and epoxy resin. 
    (b) Single-diameter sapphire powder samples. 
    (c) PTFE, HDPE (transparent and black), Epoxy UHU+ Endfest 300 and Stycast 2850FT. (d) Eccosorb CR124 and epoxy resin mixed with metal powders (copper and stainless steel).}
\end{figure}

To assess the IR blocking regime, we measured the infrared absorption of electromagnetic radiation with the wavelength between \SI{1}{\micro\meter} and \SI{1000}{\micro\meter} for various materials (Table~\ref{tab:transmission}) using a commercial IR spectrometer. The supplementary materials provide details on the measurement setup. Unless otherwise noted, the samples have a thickness of 1.5\,mm\cite{griedelSupplementaryMaterials2025}. The absorption is compared in Fig.~\ref{fig:IR_absorption}, please note the different scales. At wavelengths below \SI{200}{\micro\meter}, the SP0.45-80 and SP0.45-700 compound are nearly fully absorbing, dominated by the \SI{0.45}{\micro\meter} to \SI{80}{\micro\meter} spheres. In the FIR range (\SI{200}{\micro\meter} to \SI{1000}{\micro\meter}), absorption decreases, consistent with the Rayleigh scattering limit for wavelengths larger than the largest grain size in the mixtures. The higher absorption of SP0.45-80 compared to SP0.45-700 will be discussed below.

Sample compounds consisting of single-diameter spheres (Fig.~\ref{fig:IR_absorption}(b)), also show a high absorption. The \SI{170}{\micro\meter} sample approaches the Rayleigh scattering limit at $\lambda \approx 500\,\mathrm\mu$m, resulting in reduced absorption. This limit is expected for spheres with diameters of \SI{340}{\micro\meter} and \SI{700}{\micro\meter}, at wavelength of approximately \SI{1070}{\micro\meter} and \SI{2200}{\micro\meter}, respectively (see Fig.~\ref{fig:introduction}(d)).

The absorption of PTFE, HDPE, Stycast 2850FT, Eccosorb CR124 and UHU plus Endfest 300 samples mixed with copper as well as stainless steel powders (mass ratio 1:2) are shown in  Fig.~\ref{fig:IR_absorption}(c,d). 

HDPE (transparent) and PTFE, shows very low absorption, predominantly in the FIR region, demonstrating that they are indeed transparent to most thermal infrared radiation. For HDPE (black) (likely with a carbon filler), radiation is nearly completely blocked in the low MIR range (between \SI{1}{\micro\meter} and \SI{200}{\micro\meter}) and increased towards the FIR range.  Stycast 2850FT and epoxy resin show a high MIR absorption, with an increasing transmission at higher wavelengths. 
The metal powder samples and Eccosorb CR124 exhibit the highest absorption of the tested materials in the full IR range, extending even to the microwave regime, see the discussion below. They do not reach the typical asymptotic Rayleigh limit.

\begin{figure}
    \centering
    \includegraphics[width=0.48\textwidth]
        {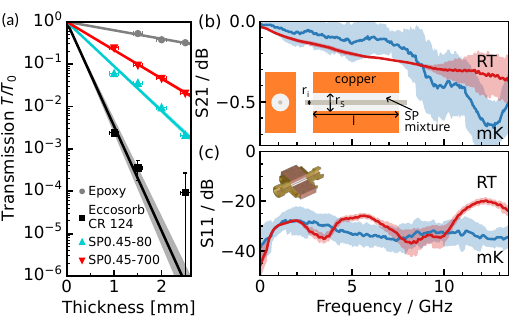}
    \caption{\label{fig:microwave_spectroscopy_filter}(a) Transmission as a function of thickness at \SI{864}{\micro\meter} for the SP mixtures, epoxy resin, and Eccosorb CR124. The solid line is a fit based on the Beer–Lambert law.  (b,c) VNA transmission and reflection measurements of an filter implementation based on the SP0.45-700 compound. The insets show the schematics (r$_i=0.4$\,mm, r$_s=2.2$\,mm, $l=8$\,mm). Red lines represent the measurements at room temperature (RT), blue at 15\,mK.}
\end{figure}

By measuring the samples at different thicknesses (\SI{1}{\milli\meter}, \SI{2}{\milli\meter}, \SI{2.5}{\milli\meter}), we can fit the Beer–Lambert law obtain the wavelength-dependent extinction coefficient $\mu_\mathrm{ext}$ for each material\cite{griedelSupplementaryMaterials2025}. In Fig.~\ref{fig:microwave_spectroscopy_filter}(a), this is  shown for a exemplary wavelength of \SI{864}{\micro\meter} for the SP mixtures, epoxy resin, and Eccosorb CR124. 
 
\begin{table*}[bt]
\centering
\caption{\label{tab:transmission} Overview of the transmission at selected wavelength of the investigated materials and a thickness of 1.5\,mm.
}

\sisetup{
        table-align-uncertainty = true,
        }
\begin{ruledtabular}
\begin{tabular*}{\textwidth}{@{\extracolsep{\fill}}l 
    S[table-format=<1.2(1)e-1]
    S[table-format=<1.2(1)e-1]
    S[table-format=<1.2(1)e-1]
    S[table-format=<1.2(1)e-1]
    S[table-format=<1.2(1)e-1] }
Material& \text{$\lambda = 2\,\mu$m} & \text{$\lambda = 40\,\mu$m} & \text{$\lambda = 200\,\mu$m} & \text{$\lambda = 500\,\mu$m} & \text{$\lambda = 700\,\mu$m} \\
\hline
SP0.45-80   & \num{3.8(3.3)e-4}  & \num{<8.1e-4}     & \num{1.1(7)e-4}   & \num{5.7(7)e-4}   & \num{6.8(7)e-3}    \\
SP0.45-700  & \num{<4.4e-5}      & \num{<8.1e-4}     & \num{<5.3e-5}     & \num{8.7(6)e-3}   & \num{3.24(23)e-2}  \\
SP180\,$\mu$m & \num{1.4(5)e-3}    & \num{<8.1e-4}     & \num{<5.3e-5}     & \num{<5.7e-5}     & \num{1.18(13)e-2}  \\
SP340\,$\mu$m & \num{3.1(4)e-3}    & \num{<8.1e-4}     & \num{<5.3e-5}     & \num{2.7(1.8)e-4} & \num{5.2(8)e-4}    \\
SP700\,$\mu$m & \num{7.5(5)e-3}    & \num{<8.1e-4}     & \num{1.58(52)e-4} & \num{2.47(18)e-3} & \num{5.13(11)e-3}  \\
PTFE          & \num{4.63(5)e-2}   & \num{2.09(29)e-2} & \num{7.5(8)e-1}   & \num{8.90(24)e-1} & \num{8.29(22)e-1}  \\
HDPE (transparent)     & \num{3.96(5)e-1}   & \num{8.23(8)e-1}  & \num{8.84(18)e-1} & \num{8.87(21)e-1} & \num{8.98(25)e-1}  \\
HDPE (black)     & \num{<4.4e-5}      & \num{<8.1e-4}     & \num{1.67(2)e-1}  & \num{4.92(11)e-1} & \num{6.06(17)e-1}  \\
UHU plus Endfest 300 & \num{2.89(11)e-2}  & \num{8.2(7.6)e-4} & \num{5.15(14)e-3} & \num{1.72(7)e-1}  & \num{3.32(13)e-1}  \\
Stycast 2850FT& \num{<4.4e-5}     & \num{<8.1e-4}     & \num{1.51(10)e-3} & \num{9.9(4)e-2}   & \num{1.93(8)e-1}   \\
Eccosorb CR124& \num{<4.4e-5}     & \num{<8.1e-4}     & \num{<5.3e-5}     & \num{<5.7e-5}     & \num{8.4(1.6)e-4}  \\
Copper powder & \num{<4.4e-5}     & \num{<8.1e-4}     & \num{<5.3e-5}     & \num{<5.7e-5}     & \num{<1.0e-4}  \\
Stainless steel powder& \num{<4.4e-5} & \num{<8.1e-4} & \num{<5.3e-5}     & \num{9.9(1.6)e-4} & \num{3.2(4)e-2}    \\
\hline
Detector limit& \num{4.4e-5}      & \num{8.1e-4}      & \num{5.3e-5}      & \num{5.7e-5}      & \num{1.0e-4}    \\
\end{tabular*}
\end{ruledtabular}
\end{table*}

We expect no significant change at cryogenic temperatures for the SP material in the infrared range, as the strong interaction of Mie scattering is only weakly temperature dependent and arises from changes in the refractive index, thermal expansion and the material’s polarizability (order 10\,\%), visible predominately around the resonant wavelength near \SI{25}{\micro\meter}\cite{querryOpticalConstants1985,halpernFarInfraredTransmission1986}. The situation in the microwave regime, however, is temperature-dependent due to the material's thermal contraction and the changed dielectric constant. 

To test this, filter prototypes were designed and built specifically for the SP0.45-700 compound, as illustrated in Fig.~\ref{fig:microwave_spectroscopy_filter}. We optimized the dimensions to match the 50\,Ohm impedance of the microwave network, particularly for in Kelvin and millikelvin temperature ranges\cite{griedelSupplementaryMaterials2025}. This can be seen in Fig.~\ref{fig:microwave_spectroscopy_filter}(b,c), where the reflection of cold samples shows a much flatter response around $-30\,$dB in the range of 0--14\,GHz and lower attenuation, e.g., at 5\,GHz of 0.1\,dB (mK) vs 0.2\,dB (RT), with only small variations between different samples. The extracted filter impedances are 52.8\,Ohm (RT) and 51.9\,Ohm (mK) at 10 GHz deviate slightly from the target value but are sufficient for quantum applications. From this, we exptract a small dielectric constant change from $\epsilon_r=3.75$ (RT) to $\epsilon_r=3.88$ (mK).

\begin{figure}
    \centering
    \includegraphics[width=0.48\textwidth]{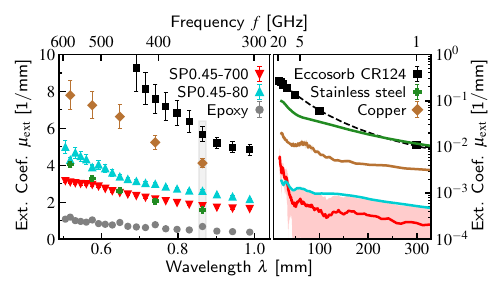}
    \caption{\label{fig:extinction_combined}
    The extinction coefficients of SP, epoxy resin, and Eccosorb CR124 are shown as a function of wavelength. The left graph shows the infrared range (gray region
    indicates the exemplary fit at \SI{864}{\micro\meter} in
    Fig.~\ref{fig:microwave_spectroscopy_filter}(a)). The right graph
    shows microwave data of the SP compound based filter shown in
    Fig.~\ref{fig:microwave_spectroscopy_filter}(b,c) as well as for
    copper and stainless steel. The Eccosorb CR124 data are taken
    from Ref.~\onlinecite{lairdEccosorbrMFDatasheetEccosorb2015}. Note the
    different y-scale on the left and right graph.}
\end{figure}
Figure~\ref{fig:extinction_combined} revisits the initial requirements of the low-pass filter. The critical FIR region, which is below $\lambda = 1.0$\,mm (the limit of the IR setup), is compared with the transmission at microwave frequencies. Materials such as copper powder, stainless steel powder, or Eccosorb CR124 exhibit the anticipated high absorption, approximately 2 times higher than the SP compounds.
Based on the measurement data, we estimate a lower bound of absorption for the SP compounds of $1.2\times10^{-6}$ (118 dB) for $l=8$\,mm and $\lambda = 1,0$\,mm. We note that the extinction data for the SP0.45-700 compound is lower than that of the SP0.45-80 compound due to a finite-size effect in the IR absorption measurement: The 170\,$\mu$m, 340\,$\mu$m, 700\,$\mu$m sapphire spheres occupy a large volume within the 1–2.5\,mm thick samples, leading to a low number of scattering events (see Fig.~\ref{fig:microwave_spectroscopy_filter}(a)). In a real filter with $l\gg 2.5$\,mm, the number of Mie scattering events is substantially higher, making it more efficient. 

In the pass-band at 9\,GHz ($\lambda = 33.3$\,mm), as shown in Fig.~\ref{fig:extinction_combined}(right), we find that the extinction coefficient is approximately $\mu_{\mathrm{ext}}\approx 1.9\times10^{-3}$\,/mm for the SP0.45-700 compound and approximately $1.9\times10^{-1}$\,/\,mm for Eccosorb CR124.

In summary, we propose a non-magnetic material compound consisting of epoxy resin and sapphire spheres of adapted sizes to achieve a high infrared attenuation (stop-band) and minimal attenuation in the microwave range (pass-band). We simulated the composition using Mie scattering theory, which is useful for estimating the extinction length in the case of strong photon-sphere interaction. The simulation predictions were tested using infrared absorption measurements on the material system, and compared to several conventional materials used in low-temperature quantum applications.  The experimental results confirm the desired low-pass behavior, demonstrating absorption exceeding $\mu_{\mathrm{ext}}\approx 2$\,/mm up to far-infrared wavelengths and $\mu_{\mathrm{ext}}\approx 4\times10^{-4}$\,/mm in the GHz regime. A low-pass filter prototype made from the SP0.45-700 compound was tested at millikelvin temperatures. Compared with measured metallic powders  of copper and stainless steel, SP0.45‑700 shows a similar stop‑band attenuation and about $25\,\times $ lower pass‑band attenuation up to 10\,GHz.

\section*{Supplementary Material}
See the supplementary material for additional details on Mie
scattering simulations (A), the filter material and sample preparation
(B),
the IR measurement and data processing (C),  microwave measurement
details (D) and additional IR measurement
data (E). It includes Refs.~\onlinecite{bohrenAbsorptionScatteringLight1998,finaladvancedmaterials2MS001FINALAdvanced2024,querryOpticalConstants1985,uhuEndfest300Datenblatt2025,savitzkySmoothingDifferentiationData1964}.

\begin{acknowledgments}
 We acknowledge technical lab support by Sebastian Koch and Rebecca Zwickel. This research was supported by the German Federal Ministry of Education and Research under the Research Program Quantum Systems, through the projects GeQCoS (FZK13N15691), qBriqs (FZK13N15950) and Qrious (FZK13N17125) as well as by funding from the European Research Council (ERC Advanced Grant {\em Milli-Q}, GAN101054327).

\end{acknowledgments}

\section*{Data Availability Statement}
The data that support the findings of
this study are available from the
corresponding authors upon reasonable
request.

\section*{Author Contributions}
M.G., M.K. and H.R. fabricated the samples, performed the measurements and analyzed
the data. B.G. and Y.L.M contributed to the IR-measurements. M.G. and
H.R. developed the theoretical model and wrote the manuscript with
input from all authors. A.V.U. and H.R. provided the experimental
means and supervised the project.

\bibliography{IR_filter_paper}

\newpage
\newpage
\onecolumngrid
\setcounter{table}{0}
\renewcommand{\thetable}{S\arabic{table}}
\setcounter{figure}{0}
\renewcommand{\thefigure}{S\arabic{figure}}
\renewcommand{\thesection}{\Alph{section}}
\renewcommand{\theequation}{S.\arabic{equation}}

\newpage
\section{Mie Scattering Simulation Background}

The scattering efficiency is calculated using the experimentally measured complex refractive index\cite{querryOpticalConstants1985} up to 60\,$\mu$m. Beyond this wavelength, the refractive index is assumed constant, as no additional vibrational modes are expected in this spectral region. The polymer matrix with real refractive index $n_{\text{env}}$ is included as the surrounding dielectric medium in the simulations, which modifies the refractive index according to
\begin{equation}
   m_\mathrm{eff} = \frac{n - i\kappa}{n_{\text{env}}}.
\end{equation}
Mie theory provides an exact solution to Maxwell’s equations, accounting for the full angular and wavelength dependence of the scattered field. The only required input parameters are the size parameter \(x\) and the particle’s relative complex refractive index \(m\). Following Ref.~\onlinecite{bohrenAbsorptionScatteringLight1998}, the extinction cross-section is given by

\begin{equation}
C_{\text{ext}} = 2 \pi \lambda_\mathrm{eff}^2 \sum_{n=1}^{\infty} 
    (2n + 1)\, \mathrm{Re} (a_n + b_n),
\label{eq:mie_scatteringxsection}
\end{equation}

where \(\lambda_\mathrm{eff}\) denotes the effective wavelength in the surrounding medium, and \(a_n\) and \(b_n\) are the Mie coefficients. These are defined as

\begin{equation}
a_n = \frac{m \psi_n(mx) \psi_n'(x) - \psi_n(x) \psi_n'(mx)}{m \psi_n(mx) \xi_n'(x) - \xi_n(x) \psi_n'(mx)},
\end{equation}

\begin{equation}
b_n = \frac{\psi_n(mx) \psi_n'(x) - m \psi_n(x) \psi_n'(mx)}{\psi_n(mx) \xi_n'(x) - m \xi_n(x) \psi_n'(mx)},
\end{equation}

where \(\psi_n\) and \(\xi_n\) are the Riccati--Bessel functions, which contain the dependence on the wavelength.

At smaller wavelengths, the scattering efficiency becomes constant, while at larger wavelengths, it decays exponentially at a wavelength dependent on the diameter of the sphere. This allows for the definition of a sharp cutoff at higher frequencies, determined by the size of the largest grains, while still maintaining significant scattering efficiency at wavelengths smaller than the smallest grains in the mixture.

\section{Preparation of the Samples}
\label{apx:fabrication}
\begin{figure}
    \centering
    \includegraphics[width=0.35\textwidth]{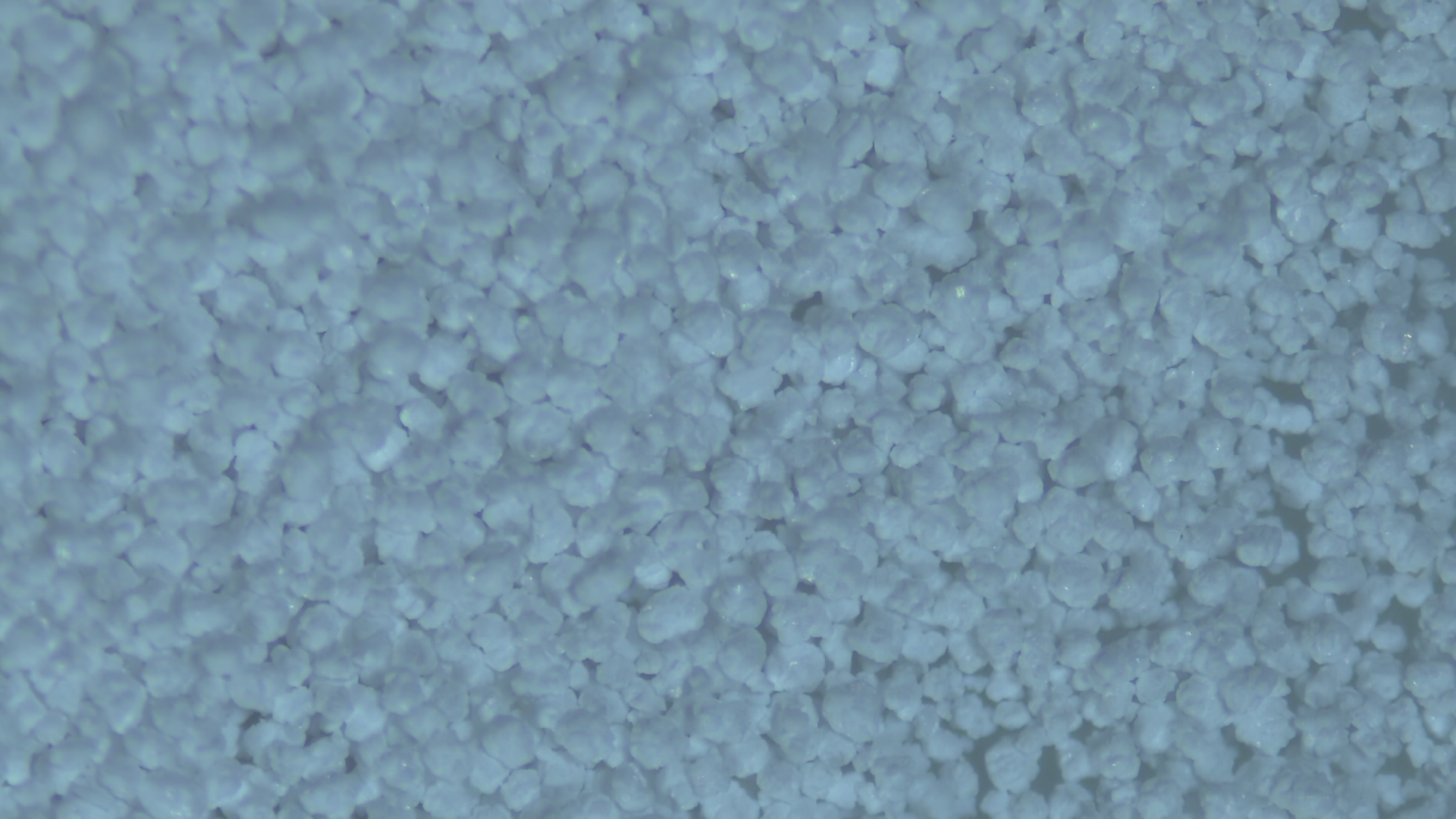}
    \includegraphics[width=0.26\textwidth]{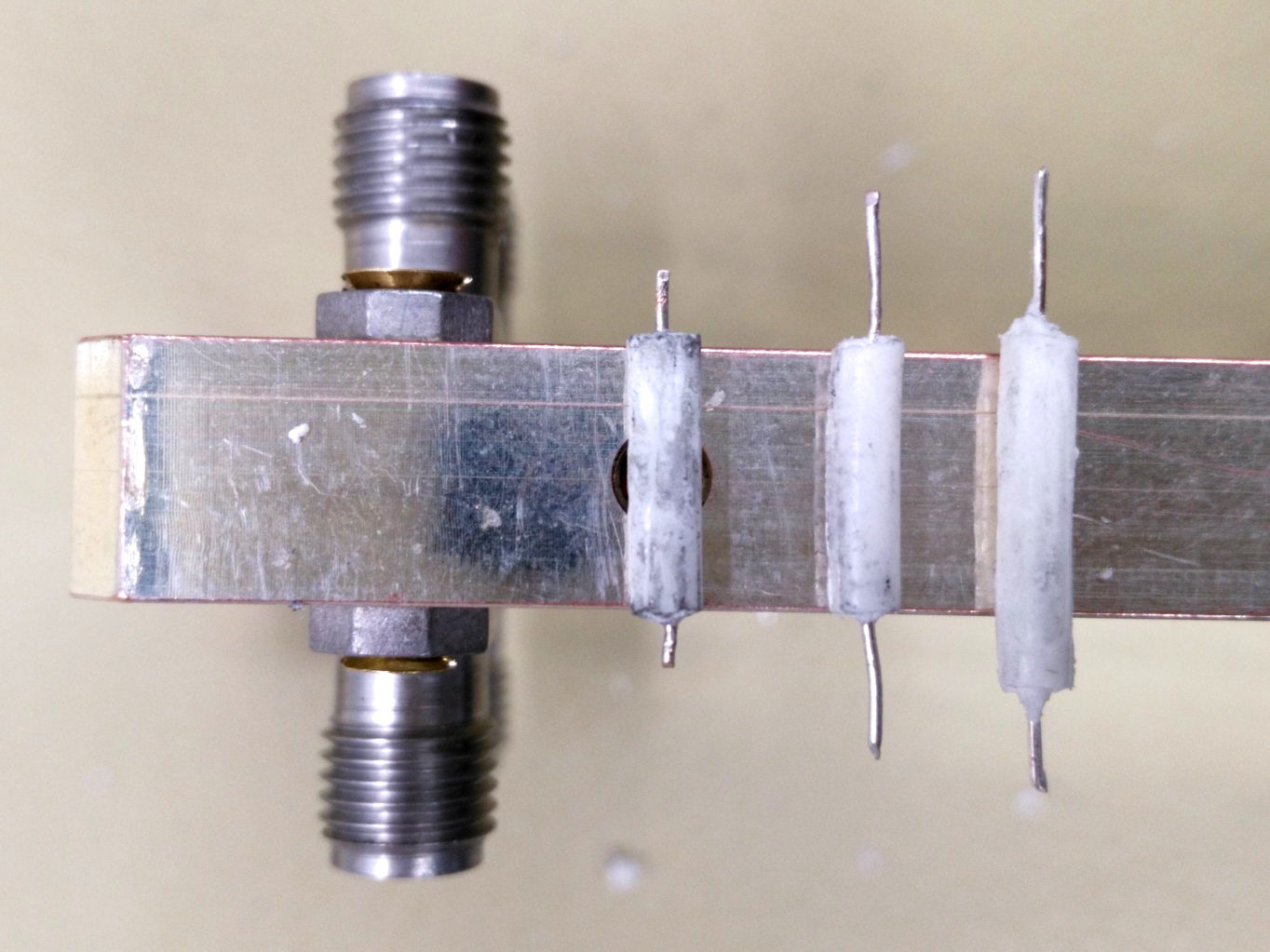}
    \caption{\label{fig:fabrication}left: Microscope picture of the powder mixture SP0.45-80. 
    \textit{right}: 8\,mm copper microwave sample with SMA connectors, variety of SP0.45-700 insets.}
\end{figure}

For the sapphire sample, spheres with diameters of $0.45$, $2.5$, $34$, $60$, $80$, $170$, $340$, and $700\,$µm were investigated\cite{finaladvancedmaterials2MS001FINALAdvanced2024}. The grains are approximately spherical, though not perfectly so. For a microscopic image of the SP0.45-80 mixture, see Fig.~\ref{fig:fabrication}, {left}. 

The individual batches of sapphire grains contain a manufacturer
selection related distribution of sizes, where the median is given
above. In the Mie calculation, we only considered the median. However,
we argue that the larger effective distribution of grain sizes is not
detrimental but rather beneficial. It smooths the extinction
efficiency spectrum shown in Fig.~1(a) of the main text.

The epoxy resin serves two requirements. First it separates the grains
in the compound, with a distinct $n_\mathrm{env}$. Second, it serves
as a binding matrix for the powder, ensuring mechanical stability and
homogeneous distribution of the grains. For preparation of the
sapphire filter samples, the two components of the UHU plus Endfest
300 adhesive (binder and hardener) were mixed in a 1:1 volume
ratio\cite{uhuEndfest300Datenblatt2025}. For composite preparation,
sapphire powder mixtures with controlled particle size distributions
were incorporated (for grading and mixing ratios, see Tab.~1 of the
main text).

Prior to curing, the mixture was vacuum-degassed for 5 minutes to eliminate entrapped air. The samples were cured under ambient conditions for 18--24 hours or heated to \SI{50}{\degree C} for quicker curing. For infrared spectrometer measurements, round disks with a diameter of \SI{13}{mm} and thickness of \SI{1}{mm}, \SI{1.5}{mm}, \SI{2}{mm} and \SI{2.5}{mm} were prepared. The metal-based mixtures, as well as the Stycast 2850FT and Eccosorb CR124 samples, were prepared in a similar manner. The HDPE and PTFE samples were cut from larger sheets.

\section{IR measurement details}
\label{apx:spectrometer}

\begin{figure}
    \includegraphics[width=0.78\textwidth]
                {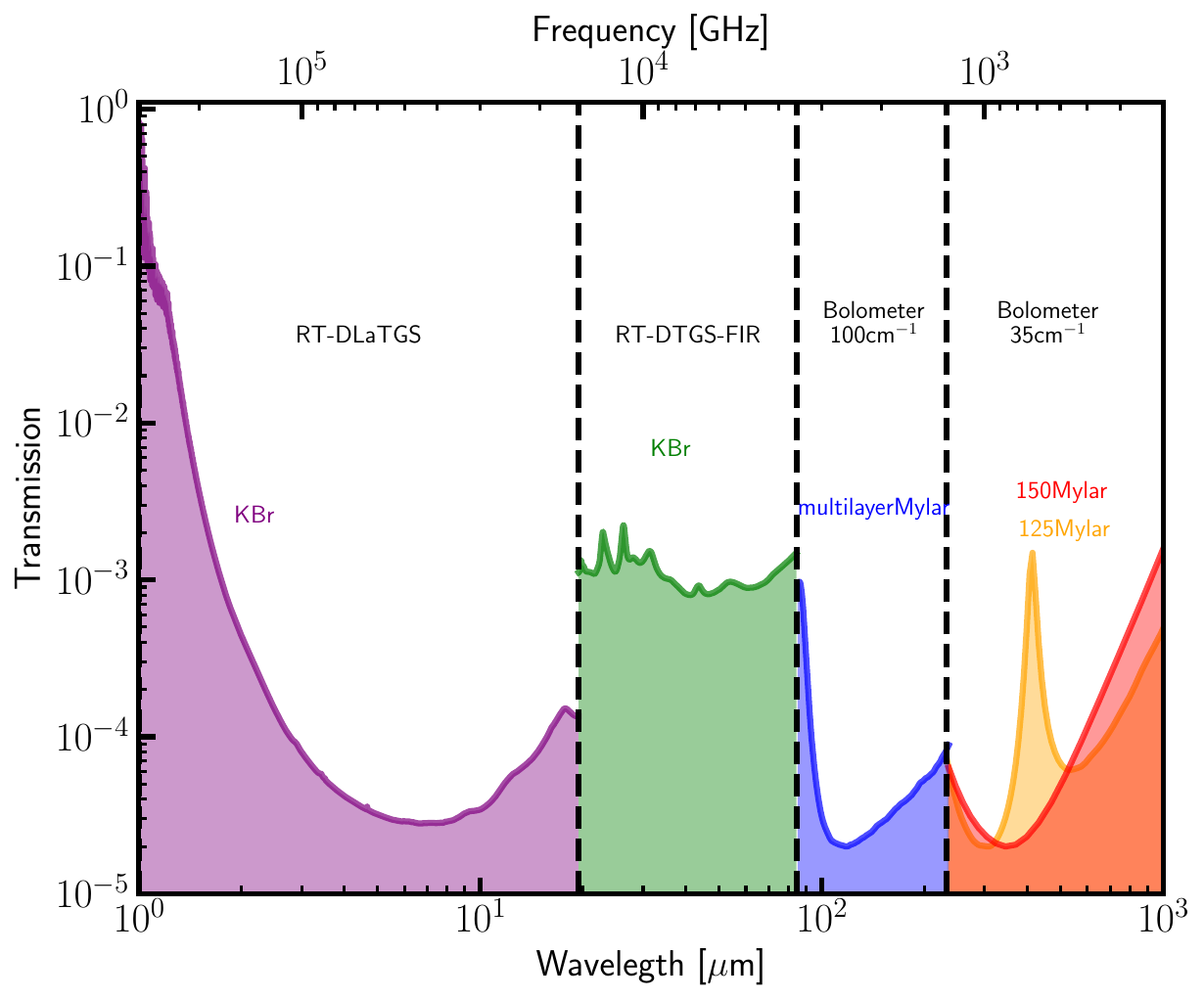}
    \caption{\label{fig:background}Background measurements providing the sensitivities of the different detector-beamsplitter combinations (filled colored areas). The wavelength limits of the different used detectors are indicated by the vertical dashed lines.}
\end{figure}

A Bruker VERTEX 80v spectrometer equipped with three detectors was used for the measurements, the radiation source for the measurements was a silicon carbide Globar. For the wavelength ranges \SIrange{1}{19.4}{\micro\meter} and \SIrange{19.4}{84.7}{\micro\meter}, DLaTGS and DTGS-FIR detectors, respectively, were employed; both are deuterated triglycine sulfate detectors. For wavelengths above \SI{84.7}{\micro\meter}, a silicon bolometer cooled to \SI{1.6}{\kelvin} was used along with two filters: one with a lower cutoff at \SI{100}{\micro\meter} (\SI{100}{\per\centi\meter}) and another at \SI{285.7}{\micro\meter} (\SI{35}{\per\centi\meter}). The beamsplitter was changed for each wavelength range: a potassium bromide beamsplitter (KBr) for \SIrange{1}{84.7}{\micro\meter}, a multilayer Mylar beamsplitter for \SIrange{84.7}{238.1}{\micro\meter}, and a \SI{150}{\micro\meter} Mylar beamsplitter for longer wavelengths. Some long-wavelength measurements utilized a \SI{125}{\micro\meter} Mylar beamsplitter. Measurements were taken with high spectral resolution between \SI{1}{\per\centi\meter} and \SI{4}{\per\centi\meter}. Depending on the expected intensity, an aperture of either \SI{4}{\milli\meter} or \SI{10}{\milli\meter} was used. All measurement settings are summarized in Tab.~\ref{tab:settings_all_measurements}.

The choice of detector and beamsplitter defines the wavelength-dependent experimental sensitivity. Calibration was performed by recording vacuum transmission for all detectors to establish the  \textit{100\%-line}. Transmission values are defined relative to this reference. Figure~\ref{fig:background} shows the detector sensitivity limit, derived from the 100\%-line. The transmission data presented in the main text, was corrected for this sensitivity.

The detector–beamsplitter combinations exhibit frequency-dependent sensitivities. The deuterated triglycine sulfate detectors show a parabolic sensitivity curve, with maximum sensitivities of \(3\times10^{-5}\) at \(6~\mu\text{m}\) and \(1\times10^{-3}\) at \(40~\mu\text{m}\). The bolometer's peak sensitivity is \(2\times10^{-5}\) at \(350~\mu\text{m}\). This wavelength-dependent sensitivity defines the lower bound for measurable transmission through the materials. The bolometer, combined with the 35 cm\(^{-1}\) filter and the \(100\,\mu\text{m}\) Mylar beamsplitter, displays peak sensitivity around \(400~\mu\text{m}\) and increased sensitivity toward the long-wavelength end near \SI{1000}{\micro\meter}, reflected in the measured data.

The data, recorded at resolutions between 1 cm\(^{-1}\) and 4 cm\(^{-1}\), were smoothed using a third-order polynomial Savitzky-Golay filter~\cite{savitzkySmoothingDifferentiationData1964}. The window length varied from 150 points for the DLaTGS detector to 30 points for the DTGS-FIR and the 1.6 K bolometer. The reported measurement error arises mainly from the variance within the smoothing window. Additional uncertainties arise from calibration variations (notably in bright measurements). 

From the spectrum we calculated the absorption and transmission data, where different wavelength ranges were combined. Measurement data for samples of all thicknesses are presented in the main text and in Sec.~\ref{ch:further_results}.

\section{Background of the Microwave Measurements}

The cryogenic measurements were performed in a \textit{Bluefors LD250} dilution refrigerator at a temperature of 15\,mK. The samples were mounted at the base plate. Microwave lines with a low attenuation were calibrated in several cooldowns, using additional microwave lines without samples for reference. The data was taken using a Vector Network Analyzer (\textit{Keysight E5080B}) over a frequency range of 0.1 to \SI{14}{\giga\hertz} at high photon numbers to maintain a sufficient signal-to-noise ratio.

\section{IR spectroscopy transmission data}
\label{ch:further_results}

\begin{figure}
    \centering
    \includegraphics[width=0.7\textwidth]{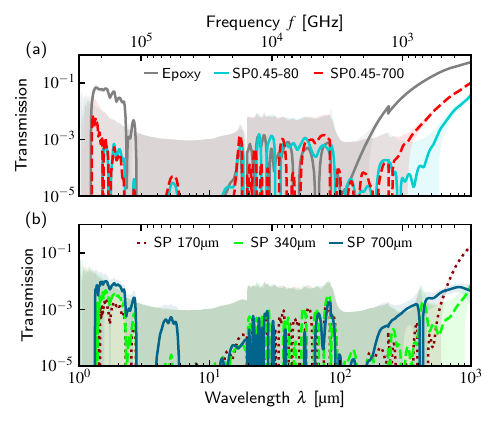}
    \caption{\label{fig:sapphire}Transmission of SP mixed with epoxy resin with a thickness of $1.5\,$mm. Shaded area represents the error. a: Transmission of the two SP mixtures (SP0.45-80: light blue, SP0.45-700: red) as well as of epoxy resin (gray) b: Transmission through single sized spheres (170\,µm: blue, 340\,µm: green, 700\,µm: dark red).}
    
\end{figure}

In Fig.~\ref{fig:sapphire}(a), the transmission is shown for the two mixtures of sapphire spheres in comparison with the transmission of epoxy resin.

In the region between \SI{1}{\micro\meter} and \SI{200}{\micro\meter}, referred to as the mid-infrared (MIR) range, the epoxy resin sample exhibits partial transparency at short wavelengths, whereas the sapphire-sphere mixtures show very low transmission within the error bars of the detector's sensitivity limit. According to the simulations discussed above, this is due to the presence of small spheres in the mixtures with diameters of \SI{0.45}{\micro\meter} and \SI{2.5}{\micro\meter}.

In the range between \SI{20}{\micro\meter} and \SI{100}{\micro\meter}, all samples show lower average transmission accompanied by larger uncertainties. This behavior arises from the use of a different detector in this spectral region, which has lower sensitivity. The measured transmission of the samples is at or below the detector’s sensitivity limit of about \num{1e-4} on average. A potentially higher transmission in this region, possibly caused by missing spheres with diameters between \SI{2.5}{\micro\meter} and \SI{34}{\micro\meter}, cannot be resolved. Therefore, the sapphire-sphere mixtures exhibit nearly perfect absorption in the MIR range.

\begin{figure}
    \centering
    \includegraphics[width=0.65\textwidth]{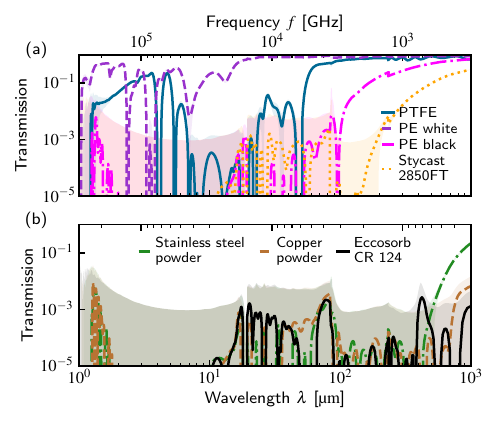}
    \caption{\label{fig:IR_spectroscopy_materials}Transmission of additionally investigated materials as function of wavelength. Shaded area represents the error. a: Commercial materials: PTFE, HDPE (transparent), HDPE (black). b: Eccosorb CR124, Stycast 2850FT and metal powder samples (copper and  stainless steel).}
\end{figure}

In Fig.~\ref{fig:sapphire}(b), the transmission through samples
consisting of single-sized spheres is shown. In the MIR range, these
sapphire samples exhibit a double peak near \SI{2}{\micro\meter},
which coincides with the peak observed in the epoxy resin sample. This
is expected since there are no small-diameter particles present with
significant extinction in this spectral range. The large spheres leave
considerable space between them through which short wavelengths can
pass, as there are fewer particles within the same volume. As shown in
Eq.~1 in the main text, the extinction coefficient depends on the total number
density. For the \SI{700}{\micro\meter} spheres, a second smaller peak
near \SI{4}{\micro\meter} is observed; its origin is unclear. We
speculate that it arises from a path located entirely inside the
sapphire, since the transmission measurement of the epoxy resin sample
shows no passband at these wavelengths. Such a path becomes more
probable for larger sphere diameters, as only a few spheres in direct
contact are needed. Given a sample thickness of \SI{1.5 \pm
0.2}{\milli\meter} and a sphere size of \SI{0.7 \pm
0.2}{\milli\meter}, only two to three spheres must be in direct
contact. This hypothesis is supported by measurements on samples of
varying thickness: for thicker samples (\SI{2}{\milli\meter} and
\SI{2.5}{\milli\meter}), the peak disappears, while for thinner
samples (\SI{1}{\milli\meter}) it also appears for smaller spheres
(see Fig.~\ref{fig:IR_spectroscopy_materials_supplementary}). In the
FIR range, the transmission of the \SI{170}{\micro\meter} spheres
shows a dip around \SI{500}{\micro\meter}, corresponding to the
wavelength of maximal extinction predicted by simulation.
Figure~\ref{fig:IR_spectroscopy_materials} shows the transmission of
the other materials for comparison.

\begin{figure}
    \centering
    \includegraphics[width=0.65\textwidth]{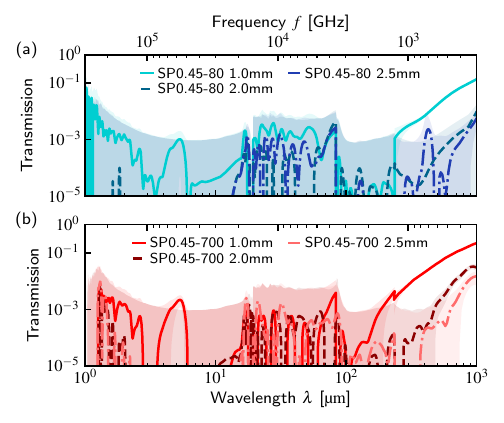}
    \caption{\label{fig:thickness}Comparison of different thicknesses of samples: 1.0\,mm (solid line), 2.0\,mm (dashed line) and 2.5\,mm (dot-dashed line). The errors are indicated by the shaded area. a: SP0.45-80, b: SP0.45-700. 
    }
    
\end{figure}

\begin{figure}
    \centering
    \includegraphics[width=0.46\textwidth]{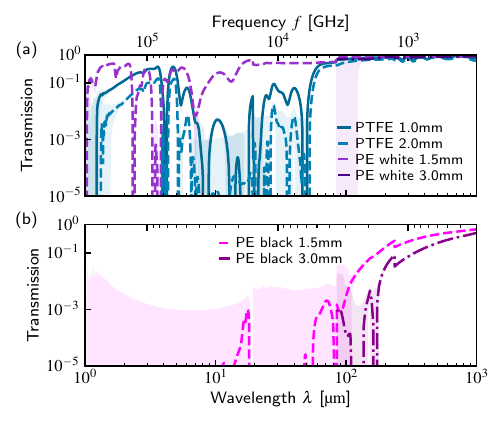}
    \includegraphics[width=0.46\textwidth]{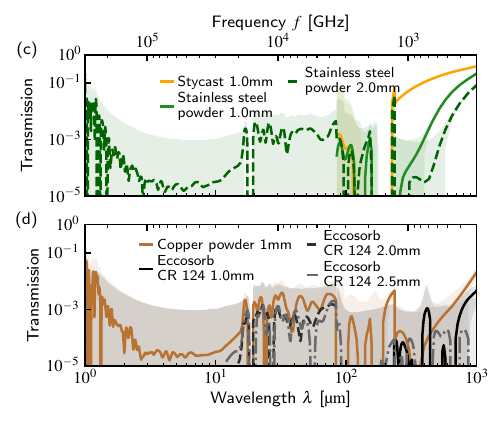}
    \caption{\label{fig:IR_spectroscopy_materials_supplementary}Transmission of the investigated materials as function of wavelength. Shaded area represents the error. a: Commercial materials: PTFE and HDPE (transparent). b: HDPE (black). c: Stycast 2850FT and stainless steel powder sample. d: copper powder sample and Eccosorb CR124.}
\end{figure}

\begin{figure}
    \centering
\includegraphics[width=0.46\textwidth]{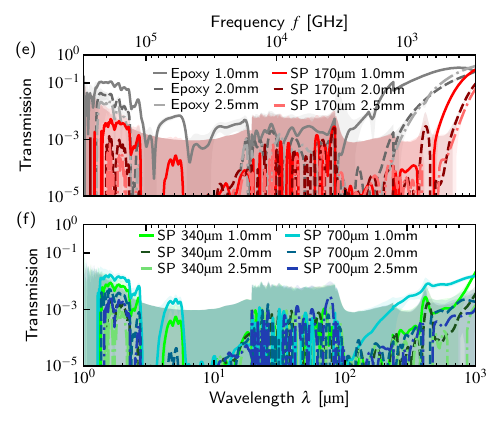}
 \caption{ \label{fig:IR_spectroscopy_materials_supplementary2}Transmission of the investigated materials (continued). e: UHU plus Endfest 300 and SP 170µm. f: SP 340µm and SP 700µm.}
\end{figure}

Figure~\ref{fig:thickness} shows the transmission through different thicknesses of samples from the two sapphire mixtures. The overall transmission decreases with increasing thickness. In most regions of the MIR regime, the transmission remains at the detector's sensitivity limit, aside from features similar to those observed in the \SI{1.5}{\milli\meter} samples discussed above. In Fig.~\ref{fig:IR_spectroscopy_materials_supplementary} and~\ref{fig:IR_spectroscopy_materials_supplementary2}, the transmission of the other materials at additional thicknesses is shown; the same decreasing transmission with thickness can be seen. Table~\ref{tab:full_transmission} provides an overview of all transmissions of all measured samples. 

\begin{table}
    \centering
    \caption{\label{tab:full_transmission}Additional data for the 1\,mm, 2\,mm and 2.5\,mm samples. See also Fig.~\ref{fig:IR_spectroscopy_materials_supplementary}. }
    \begin{ruledtabular}
    \begin{tabular}{l c c c c c c}
    Material & Thickness & 2~\textmu m & 40~\textmu m & 200~\textmu m & 500~\textmu m & 700~\textmu m \\
             & (mm) & \multicolumn{5}{c}{Transmission} \\
    \hline
    UHU plus Endfest 300 & 1.0  & \num{6.69(16)e-2} & \num{2.6(1.1)e-3} & \num{4.47(44)e-2} & \num{2.71(21)e-1} & \num{3.50(26)e-1} \\
    UHU plus Endfest 300 & 1.5 & \num{2.89(11)e-2} & \num{8.2(7.6)e-4} & \num{5.15(14)e-3} & \num{1.722(68)e-1} & \num{3.32(13)e-1} \\
    UHU plus Endfest 300 & 2.0  & \num{1.440(77)e-2} & \num{1.14(63)e-3} & \num{2.07(69)e-4} & \num{5.98(63)e-2} & \num{1.45(17)e-1} \\
    UHU plus Endfest 300 & 2.5 & \num{6.24(54)e-3} & \num{3.7(6.4)e-4} & \num{4.3(5.2)e-5} & \num{7.12(39)e-2} & \num{2.01(10)e-1} \\
    \hline
    PTFE & 1.0& \num{1.214(12)e-1} & \num{6.63(72)e-2} & \num{7.95(68)e-1} & \num{9.06(24)e-1} & \num{9.05(34)e-1} \\
    PTFE & 1.5& \num{4.632(51)e-2} & \num{2.09(29)e-2} & \num{7.55(76)e-1} & \num{8.90(24)e-1} & \num{8.29(22)e-1} \\
    PTFE & 2.0  & \num{2.190(39)e-2} & \num{6.52(71)e-3} & \num{6.84(69)e-1} & \num{8.62(24)e-1} & \num{7.79(14)e-1} \\
    \hline
    HDPE (transparent) & 1.5  & \num{3.966(45)e-1} & \num{8.234(84)e-1} & \num{8.84(18)e-1} & \num{8.87(21)e-1} & \num{8.98(25)e-1} \\
    HDPE (transparent) & 3.0  & \num{3.985(46)e-1} & \num{5.109(55)e-1} & \num{8.55(27)e-1} & \num{8.707(87)e-1} & \num{8.769(88)e-1} \\
    \hline
    HDPE (black) & 1.5 & \num{<1e-5} & \num{<1e-5} & \num{1.677(22)e-1} & \num{4.92(11)e-1} & \num{6.06(17)e-1} \\
    HDPE (black) & 3.0 & \num{<1e-5} & \num{<1e-5} & \num{1.58(10)e-1} & \num{4.64(21)e-1} & \num{5.79(26)e-1} \\
    \hline
    Stycast 2850 FT, & 1.0 & \num{<1e-5} & \num{<1e-5} & \num{<1e-5} & \num{1.76(14)e-1} & \num{2.83(22)e-1} \\
    Stycast 2850 FT, & 1.5 & \num{<1e-5} & \num{<1e-5} & \num{1.509(95)e-3} & \num{9.98(38)e-2} & \num{1.930(75)e-1} \\
    \hline
    Eccosorb CR124 & 1.0 & \num{<1e-5} & \num{<1e-5} & \num{<1e-5} & \num{<1e-5} & \num{9.3(6.7)e-5} \\
    Eccosorb CR124 & 1.5 & \num{<1e-5} & \num{<1e-5} & \num{<1e-5} & \num{<1e-5} & \num{8.4(1.6)e-4} \\
    Eccosorb CR124 & 2.0 & \num{<1e-5} & \num{6.5(6.0)e-4} & \num{1.3(1.4)e-3} & \num{1.3(1.4)e-3} & \num{1.3(1.4)e-3} \\
    Eccosorb CR124 & 2.5 & \num{<1e-5} & \num{7.4(9.4)e-4} & \num{<1e-5} & \num{1.59(94)e-4} & \num{<1e-5} \\
    \hline
    Epoxy + copper& 1.0 & \num{<1e-5} & \num{<1e-5} & \num{1.03(76)e-3} & \num{<1e-5} & \num{2.86(22)e-3} \\
    Epoxy + copper  & 1.5 & \num{<1e-5} & \num{<1e-5} & \num{5.0(11)e-5} & \num{<1e-5} & \num{<1e-5} \\
    \hline
    Epoxy + stainless steel & 1.0 & \num{<1e-5} & \num{<1e-5} & \num{<1e-5} & \num{1.227(45)e-2} & \num{8.5(1.1)e-2} \\
    Epoxy + stainless steel & 1.5 & \num{<1e-5} & \num{<1e-5} & \num{<1e-5} & \num{9.9(1.6)e-4} & \num{3.23(38)e-2} \\
    Epoxy + stainless steel & 2.0 & \num{1.6(4.3)e-4} & \num{1.6(1.9)e-3} & \num{<1e-5} & \num{5.9(1.6)e-4} & \num{1.868(54)e-2} \\
    Epoxy + stainless steel & 2.5 & & & & & \\
    \hline
    SP0.45-80 & 1.0 & \num{1.52(56)e-3} & \num{1.3(1.6)e-3} & \num{2.9(6.0)e-5} & \num{1.99(18)e-2} & \num{6.58(76)e-2} \\
    SP0.45-80 & 1.5 & \num{3.8(3.3)e-4} & \num{<1e-5} & \num{1.06(66)e-4} & \num{5.73(67)e-4} & \num{6.87(72)e-3} \\
    SP0.45-80 & 2.0 & \num{<1e-5} & \num{2.0(6.8)e-4} & \num{<1e-5} & \num{3.5(1.1)e-4} & \num{1.74(20)e-3} \\
    SP0.45-80 & 2.5 & \num{<1e-5} & \num{<1e-5} & \num{<1e-5} & \num{<1e-5} & \num{5.1(11)e-4} \\
    \hline
    SP0.45-700 & 1.0 & \num{1.54(36)e-3} & \num{2.9(7.1)e-4} & \num{1.62(12)e-3} & \num{4.38(17)e-2} & \num{1.043(52)e-1} \\
    SP0.45-700 & 1.5 & \num{2.6(3.4)e-4} & \num{<1e-5} & \num{<1e-5} & \num{8.76(61)e-3} & \num{3.24(23)e-2} \\
    SP0.45-700 & 2.0 & \num{1.2(3.3)e-4} & \num{1.2(4.9)e-4} & \num{<1e-5} & \num{1.80(25)e-3} & \num{1.13(14)e-2} \\
    SP0.45-700 & 2.5 & \num{<1e-5} & \num{1.1(7.0)e-4} & \num{3.0(8.5)e-5} & \num{3.53(62)e-4} & \num{1.88(90)e-3} \\
    \hline
    SP180\,$\mu$m  & 1.0 & \num{3.82(34)e-3} & \num{<1e-5} & \num{<1e-5} & \num{1.48(37)e-3} & \num{5.14(56)e-2} \\
    SP180\,$\mu$m  & 1.5 & \num{1.45(46)e-3} & \num{<1e-5} & \num{<1e-5} & \num{<1e-5} & \num{1.18(13)e-2} \\
    SP180\,$\mu$m  & 2.0 & \num{6.5(4.1)e-4} & \num{<1e-5} & \num{<1e-5} & \num{<1e-5} & \num{3.07(40)e-3} \\
    SP180\,$\mu$m  & 2.5 & \num{<1e-5} & \num{<1e-5} & \num{<1e-5} & \num{<1e-5} & \num{1.29(15)e-3} \\
    \hline
    SP340\,$\mu$m  & 1.0 & \num{6.86(40)e-3} & \num{<1e-5} & \num{<1e-5} & \num{7.70(27)e-4} & \num{1.34(18)e-3} \\
    SP340\,$\mu$m  & 1.5 & \num{3.13(38)e-3} & \num{<1e-5} & \num{<1e-5} & \num{2.7(1.8)e-4} & \num{5.27(84)e-4} \\
    SP340\,$\mu$m  & 2.0 & \num{1.52(40)e-3} & \num{<1e-5} & \num{<1e-5} & \num{<1e-5} & \num{2.27(27)e-4} \\
    SP340\,$\mu$m  & 2.5 & \num{6.7(3.8)e-4} & \num{<1e-5} & \num{<1e-5} & \num{9.6(6.0)e-5} & \num{<1e-5} \\
    \hline
    SP700\,$\mu$m  & 1.0 & \num{1.376(43)e-2} & \num{<1e-5} & \num{5.32(64)e-4} & \num{7.29(21)e-3} & \num{8.92(29)e-3} \\
    SP700\,$\mu$m  & 1.5 & \num{7.51(47)e-3} & \num{<1e-5} & \num{1.58(52)e-4} & \num{2.47(18)e-3} & \num{5.13(11)e-3} \\
    SP700\,$\mu$m  & 2.0 & \num{3.67(36)e-3} & \num{<1e-5} & \num{1.5(7.9)e-5} & \num{8.19(40)e-4} & \num{1.36(19)e-3} \\
    SP700\,$\mu$m  & 2.5 & \num{1.86(40)e-3} & \num{<1e-5} & \num{4.6(4.0)e-5} & \num{3.8(1.4)e-4} & \num{6.04(85)e-4} \\
\end{tabular}
\end{ruledtabular}
\end{table}

\FloatBarrier
\pagebreak
\renewcommand{\arraystretch}{1.15} 
\begin{longtable}{lccccc}
\caption{\label{tab:settings_all_measurements}Details of the IR spectrometer conditions.} \\

\hline
\hline
Material & Thickness [mm] & Detector & Beamsplitter & Resolution [cm$^{-1}$] & Aperature [mm] \\
\hline
\endfirsthead

\hline
Material & Thickness [mm] & Detector & Beamsplitter & Resolution [cm$^{-1}$] & Aperature [mm] \\
\hline
\endhead

\hline
\multicolumn{6}{r}{{Continued on next page}} \\
\endfoot

\hline
\endlastfoot

SP180\,$\mu$m & 1.0 & 1.6K bolometer, 100\,cm$^{-1}$ & multilayerMylar & 1 & 4 \\
SP180\,$\mu$m & 1.0 & 1.6K bolometer, 35\,cm$^{-1}$ & 125Mylar & 1 & 8 \\
SP180\,$\mu$m & 1.0 & RT-DLaTGS & KBr & 2 & 8 \\
SP180\,$\mu$m & 1.0 & RT-DTGS-FIR & KBr & 2 & 8 \\
SP180\,$\mu$m & 1.5 & 1.6K bolometer, 100\,cm$^{-1}$ & multilayerMylar & 1 & 4 \\
SP180\,$\mu$m & 1.5 & 1.6K bolometer, 35\,cm$^{-1}$ & 125Mylar & 1 & 8 \\
SP180\,$\mu$m & 1.5 & RT-DLaTGS & KBr & 2 & 8 \\
SP180\,$\mu$m & 1.5 & RT-DTGS-FIR & KBr & 2 & 8 \\
SP180\,$\mu$m & 2.0 & 1.6K bolometer, 100\,cm$^{-1}$ & multilayerMylar & 1 & 4 \\
SP180\,$\mu$m & 2.0 & 1.6K bolometer, 35\,cm$^{-1}$ & 125Mylar & 1 & 8 \\
SP180\,$\mu$m & 2.0 & RT-DLaTGS & KBr & 2 & 8 \\
SP180\,$\mu$m & 2.0 & RT-DTGS-FIR & KBr & 2 & 8 \\
SP180\,$\mu$m & 2.5 & 1.6K bolometer, 100\,cm$^{-1}$ & multilayerMylar & 1 & 4 \\
SP180\,$\mu$m & 2.5 & 1.6K bolometer, 35\,cm$^{-1}$ & 125Mylar & 1 & 8 \\
SP180\,$\mu$m & 2.5 & RT-DLaTGS & KBr & 2 & 8 \\
SP180\,$\mu$m & 2.5 & RT-DTGS-FIR & KBr & 2 & 8 \\
SP180\,$\mu$m & 3.0 & 1.6K bolometer, 100\,cm$^{-1}$ & multilayerMylar & 1 & 4 \\
SP180\,$\mu$m & 3.0 & 1.6K bolometer, 35\,cm$^{-1}$ & 125Mylar & 1 & 8 \\
SP340\,$\mu$m & 1.0 & 1.6K bolometer, 100\,cm$^{-1}$ & multilayerMylar & 1 & 4 \\
SP340\,$\mu$m & 1.0 & 1.6K bolometer, 35\,cm$^{-1}$ & 125Mylar & 1 & 8 \\
SP340\,$\mu$m & 1.0 & RT-DLaTGS & KBr & 2 & 8 \\
SP340\,$\mu$m & 1.0 & RT-DTGS-FIR & KBr & 2 & 8 \\
SP340\,$\mu$m & 1.5 & 1.6K bolometer, 100\,cm$^{-1}$ & multilayerMylar & 1 & 4 \\
SP340\,$\mu$m & 1.5 & 1.6K bolometer, 35\,cm$^{-1}$ & 125Mylar & 1 & 8 \\
SP340\,$\mu$m & 1.5 & RT-DLaTGS & KBr & 2 & 8 \\
SP340\,$\mu$m & 1.5 & RT-DTGS-FIR & KBr & 2 & 8 \\
SP340\,$\mu$m & 2.0 & 1.6K bolometer, 100\,cm$^{-1}$ & multilayerMylar & 1 & 4 \\
SP340\,$\mu$m & 2.0 & 1.6K bolometer, 35\,cm$^{-1}$ & 125Mylar & 1 & 8 \\
SP340\,$\mu$m & 2.0 & RT-DLaTGS & KBr & 2 & 8 \\
SP340\,$\mu$m & 2.0 & RT-DTGS-FIR & KBr & 2 & 8 \\
SP340\,$\mu$m & 2.5 & 1.6K bolometer, 100\,cm$^{-1}$ & multilayerMylar & 1 & 4 \\
SP340\,$\mu$m & 2.5 & 1.6K bolometer, 35\,cm$^{-1}$ & 125Mylar & 1 & 8 \\
SP340\,$\mu$m & 2.5 & RT-DLaTGS & KBr & 2 & 8 \\
SP340\,$\mu$m & 2.5 & RT-DTGS-FIR & KBr & 2 & 8 \\
SP340\,$\mu$m & 3.0 & 1.6K bolometer, 100\,cm$^{-1}$ & multilayerMylar & 1 & 4 \\
SP340\,$\mu$m & 3.0 & 1.6K bolometer, 35\,cm$^{-1}$ & 125Mylar & 1 & 8 \\
SP700\,$\mu$m & 1.0 & 1.6K bolometer, 100\,cm$^{-1}$ & multilayerMylar & 1 & 4 \\
SP700\,$\mu$m & 1.0 & 1.6K bolometer, 35\,cm$^{-1}$ & 125Mylar & 1 & 8 \\
SP700\,$\mu$m & 1.0 & RT-DLaTGS & KBr & 2 & 8 \\
SP700\,$\mu$m & 1.0 & RT-DTGS-FIR & KBr & 2 & 8 \\
SP700\,$\mu$m & 1.5 & 1.6K bolometer, 100\,cm$^{-1}$ & multilayerMylar & 1 & 4 \\
SP700\,$\mu$m & 1.5 & 1.6K bolometer, 35\,cm$^{-1}$ & 125Mylar & 1 & 8 \\
SP700\,$\mu$m & 1.5 & RT-DLaTGS & KBr & 2 & 8 \\
SP700\,$\mu$m & 1.5 & RT-DTGS-FIR & KBr & 2 & 8 \\
SP700\,$\mu$m & 2.0 & 1.6K bolometer, 100\,cm$^{-1}$ & multilayerMylar & 1 & 4 \\
SP700\,$\mu$m & 2.0 & 1.6K bolometer, 35\,cm$^{-1}$ & 125Mylar & 1 & 8 \\
SP700\,$\mu$m & 2.0 & RT-DLaTGS & KBr & 2 & 8 \\
SP700\,$\mu$m & 2.0 & RT-DTGS-FIR & KBr & 2 & 8 \\
SP700\,$\mu$m & 2.5 & 1.6K bolometer, 100\,cm$^{-1}$ & multilayerMylar & 1 & 4 \\
SP700\,$\mu$m & 2.5 & 1.6K bolometer, 35\,cm$^{-1}$ & 125Mylar & 1 & 8 \\
SP700\,$\mu$m & 2.5 & RT-DLaTGS & KBr & 2 & 8 \\
SP700\,$\mu$m & 2.5 & RT-DTGS-FIR & KBr & 2 & 8 \\
SP700\,$\mu$m & 3.0 & 1.6K bolometer, 100\,cm$^{-1}$ & multilayerMylar & 1 & 4 \\
SP700\,$\mu$m & 3.0 & 1.6K bolometer, 35\,cm$^{-1}$ & 125Mylar & 1 & 8 \\
Epoxy + copper & 1.0 & 1.6K bolometer, 100\,cm$^{-1}$ & 150Mylar & 4 & 3 \\
Epoxy + copper & 1.0 & 1.6K bolometer, 35\,cm$^{-1}$ & 150Mylar & 4 & 3 \\
Epoxy + copper & 1.0 & RT-DLaTGS & KBr & 4 & 3 \\
Epoxy + copper & 1.0 & RT-DTGS-FIR & KBr & 4 & 3 \\
Epoxy + copper & 1.5 & 1.6K bolometer, 100\,cm$^{-1}$ & 150Mylar & 4 & 3 \\
Epoxy + copper & 1.5 & 1.6K bolometer, 100\,cm$^{-1}$ & multilayerMylar & 1 & 4 \\
Epoxy + copper & 1.5 & 1.6K bolometer, 35\,cm$^{-1}$ & 125Mylar & 1 & 8 \\
Epoxy + copper & 1.5 & 1.6K bolometer, 35\,cm$^{-1}$ & 150Mylar & 4 & 3 \\
Epoxy + copper & 1.5 & 1.6K bolometer, 35\,cm$^{-1}$ & 50Mylar & 1 & 3 \\
Epoxy + copper & 1.5 & RT-DLaTGS & KBr & 4 & 8 \\
Epoxy + copper & 1.5 & RT-DTGS-FIR & KBr & 4 & 8 \\
Eccosorb CR124 & 1.0 & 1.6K bolometer, 100\,cm$^{-1}$ & multilayerMylar & 1 & 4 \\
Eccosorb CR124 & 1.0 & 1.6K bolometer, 35\,cm$^{-1}$ & 125Mylar & 1 & 8 \\
Eccosorb CR124 & 1.5 & 1.6K bolometer, 100\,cm$^{-1}$ & multilayerMylar & 1 & 4 \\
Eccosorb CR124 & 1.5 & 1.6K bolometer, 35\,cm$^{-1}$ & 125Mylar & 1 & 8 \\
Eccosorb CR124 & 1.5 & 1.6K bolometer, 35\,cm$^{-1}$ & 150Mylar & 4 & 3 \\
Eccosorb CR124 & 1.5 & 1.6K bolometer, 35\,cm$^{-1}$ & 50Mylar & 1 & 3 \\
Eccosorb CR124 & 1.5 & RT-DLaTGS & KBr & 4 & 8 \\
Eccosorb CR124 & 1.5 & RT-DLaTGS & KBr & 4 & 8 \\
Eccosorb CR124 & 1.5 & RT-DTGS-FIR & KBr & 4 & 8 \\
Eccosorb CR124 & 1.5 & RT-DTGS-FIR & KBr & 4 & 8 \\
Eccosorb CR124 & 2.0 & RT-DLaTGS & KBr & 4 & 8 \\
Eccosorb CR124 & 2.0 & RT-DTGS-FIR & KBr & 4 & 8 \\
Eccosorb CR124 & 2.5 & 1.6K bolometer, 100\,cm$^{-1}$ & multilayerMylar & 1 & 4 \\
Eccosorb CR124 & 2.5 & 1.6K bolometer, 35\,cm$^{-1}$ & 125Mylar & 1 & 8 \\
Eccosorb CR124 & 2.5 & RT-DLaTGS & KBr & 4 & 8 \\
Eccosorb CR124 & 2.5 & RT-DTGS-FIR & KBr & 4 & 8 \\
SP0.45-80& 1.0 & 1.6K bolometer, 100\,cm$^{-1}$ & 150Mylar & 4 & 3 \\
SP0.45-80& 1.0 & 1.6K bolometer, 100\,cm$^{-1}$ & 150Mylar & 4 & 3 \\
SP0.45-80 & 1.0 & 1.6K bolometer, 100\,cm$^{-1}$ & 150Mylar & 4 & 3 \\
SP0.45-80 & 1.0 & 1.6K bolometer, 100\,cm$^{-1}$ & multilayerMylar & 1 & 4 \\
SP0.45-80 & 1.0 & 1.6K bolometer, 35\,cm$^{-1}$ & 125Mylar & 1 & 8 \\
SP0.45-80 & 1.0 & 1.6K bolometer, 35\,cm$^{-1}$ & 150Mylar & 4 & 3 \\
SP0.45-80 & 1.0 & 1.6K bolometer, 35\,cm$^{-1}$ & 150Mylar & 4 & 3 \\
SP0.45-80 & 1.0 & RT-DLaTGS & KBr & 4 & 3 \\
SP0.45-80 & 1.0 & RT-DTGS-FIR & KBr & 4 & 3 \\
SP0.45-80 & 1.5 & 1.6K bolometer, 100\,cm$^{-1}$ & multilayerMylar & 1 & 4 \\
SP0.45-80 & 1.5 & 1.6K bolometer, 35\,cm$^{-1}$ & 125Mylar & 1 & 8 \\
SP0.45-80 & 1.5 & 1.6K bolometer, 35\,cm$^{-1}$ & 150Mylar & 4 & 3 \\
SP0.45-80 & 1.5 & 1.6K bolometer, 35\,cm$^{-1}$ & 50Mylar & 1 & 3 \\
SP0.45-80 & 1.5 & RT-DLaTGS & KBr & 4 & 8 \\
SP0.45-80 & 1.5 & RT-DLaTGS & KBr & 4 & 8 \\
SP0.45-80 & 1.5 & RT-DTGS-FIR & KBr & 4 & 8 \\
SP0.45-80 & 1.5 & RT-DTGS-FIR & KBr & 4 & 8 \\
SP0.45-80 & 2.0 & 1.6K bolometer, 100\,cm$^{-1}$ & 150Mylar & 4 & 3 \\
SP0.45-80 & 2.0 & 1.6K bolometer, 100\,cm$^{-1}$ & multilayerMylar & 1 & 4 \\
SP0.45-80 & 2.0 & 1.6K bolometer, 35\,cm$^{-1}$ & 125Mylar & 1 & 8 \\
SP0.45-80 & 2.0 & 1.6K bolometer, 35\,cm$^{-1}$ & 150Mylar & 4 & 3 \\
SP0.45-80 & 2.0 & RT-DLaTGS & KBr & 4 & 8 \\
SP0.45-80 & 2.0 & RT-DLaTGS & KBr & 2 & 8 \\
SP0.45-80 & 2.0 & RT-DTGS-FIR & KBr & 4 & 8 \\
SP0.45-80 & 2.0 & RT-DTGS-FIR & KBr & 2 & 8 \\
SP0.45-80 & 2.5 & 1.6K bolometer, 100\,cm$^{-1}$ & multilayerMylar & 1 & 4 \\
SP0.45-80 & 2.5 & 1.6K bolometer, 35\,cm$^{-1}$ & 125Mylar & 1 & 8 \\
SP0.45-80 & 2.5 & RT-DLaTGS & KBr & 4 & 8 \\
SP0.45-80 & 2.5 & RT-DTGS-FIR & KBr & 4 & 8 \\
SP0.45-700 & 1.0 & 1.6K bolometer, 100\,cm$^{-1}$ & multilayerMylar & 1 & 4 \\
SP0.45-700 & 1.0 & 1.6K bolometer, 35\,cm$^{-1}$ & 125Mylar & 1 & 8 \\
SP0.45-700 & 1.0 & 1.6K bolometer, 35\,cm$^{-1}$ & 50Mylar & 1 & 3 \\
SP0.45-700 & 1.0 & RT-DLaTGS & KBr & 4 & 8 \\
SP0.45-700 & 1.0 & RT-DTGS-FIR & KBr & 4 & 8 \\
SP0.45-700 & 1.5 & 1.6K bolometer, 100\,cm$^{-1}$ & multilayerMylar & 1 & 4 \\
SP0.45-700 & 1.5 & 1.6K bolometer, 35\,cm$^{-1}$ & 125Mylar & 1 & 8 \\
SP0.45-700 & 1.5 & 1.6K bolometer, 35\,cm$^{-1}$ & 50Mylar & 1 & 3 \\
SP0.45-700 & 1.5 & RT-DLaTGS & KBr & 4 & 8 \\
SP0.45-700 & 1.5 & RT-DTGS-FIR & KBr & 4 & 8 \\
SP0.45-700 & 2.0 & 1.6K bolometer, 100\,cm$^{-1}$ & multilayerMylar & 1 & 4 \\
SP0.45-700 & 2.0 & 1.6K bolometer, 35\,cm$^{-1}$ & 125Mylar & 1 & 8 \\
SP0.45-700 & 2.0 & 1.6K bolometer, 35\,cm$^{-1}$ & 50Mylar & 1 & 3 \\
SP0.45-700 & 2.0 & RT-DLaTGS & KBr & 4 & 8 \\
SP0.45-700 & 2.0 & RT-DTGS-FIR & KBr & 4 & 8 \\
SP0.45-700 & 2.5 & 1.6K bolometer, 100\,cm$^{-1}$ & multilayerMylar & 1 & 4 \\
SP0.45-700 & 2.5 & 1.6K bolometer, 35\,cm$^{-1}$ & 125Mylar & 1 & 8 \\
SP0.45-700 & 2.5 & 1.6K bolometer, 35\,cm$^{-1}$ & 50Mylar & 1 & 3 \\
SP0.45-700 & 2.5 & RT-DLaTGS & KBr & 4 & 8 \\
SP0.45-700 & 2.5 & RT-DTGS-FIR & KBr & 4 & 8 \\
SP0.45-700 & 3.0 & 1.6K bolometer, 100\,cm$^{-1}$ & multilayerMylar & 1 & 4 \\
SP0.45-700 & 3.0 & 1.6K bolometer, 35\,cm$^{-1}$ & 125Mylar & 1 & 8 \\
HDPE (black) & 1.5 & 1.6K bolometer, 100\,cm$^{-1}$ & multilayerMylar & 1 & 4 \\
HDPE (black) & 1.5 & 1.6K bolometer, 35\,cm$^{-1}$ & 125Mylar & 1 & 8 \\
HDPE (black) & 1.5 & 1.6K bolometer, 35\,cm$^{-1}$ & 50Mylar & 1 & 3 \\
HDPE (black) & 1.5 & RT-DLaTGS & KBr & 4 & 8 \\
HDPE (black) & 1.5 & RT-DTGS-FIR & KBr & 4 & 8 \\
HDPE (black) & 1.5 & 1.6K bolometer, 100\,cm$^{-1}$ & 150Mylar & 4 & 3 \\
HDPE (black) & 1.5 & 1.6K bolometer, 35\,cm$^{-1}$ & 150Mylar & 4 & 3 \\
HDPE (black) & 1.5 & RT-DLaTGS & KBr & 4 & 8 \\
HDPE (black) & 1.5 & RT-DTGS-FIR & KBr & 4 & 8 \\
HDPE (black) & 3.0 & 1.6K bolometer, 100\,cm$^{-1}$ & 150Mylar & 4 & 3 \\
HDPE (black) & 3.0 & 1.6K bolometer, 35\,cm$^{-1}$ & 150Mylar & 4 & 3 \\
HDPE (transparent) & 1.5 & 1.6K bolometer, 100\,cm$^{-1}$ & multilayerMylar & 1 & 4 \\
HDPE (transparent) & 1.5 & 1.6K bolometer, 35\,cm$^{-1}$ & 50Mylar & 1 & 3 \\
HDPE (transparent) & 1.5 & RT-DLaTGS & KBr & 4 & 8 \\
HDPE (transparent) & 1.5 & RT-DTGS-FIR & KBr & 4 & 8 \\
HDPE (transparent) & 1.5 & 1.6K bolometer, 100\,cm$^{-1}$ & 150Mylar & 4 & 3 \\
HDPE (transparent) & 1.5 & 1.6K bolometer, 35\,cm$^{-1}$ & 150Mylar & 4 & 3 \\
HDPE (transparent) & 1.5 & RT-DLaTGS & KBr & 4 & 8 \\
HDPE (transparent) & 1.5 & RT-DTGS-FIR & KBr & 4 & 8 \\
HDPE (transparent) & 3.0 & 1.6K bolometer, 100\,cm$^{-1}$ & 150Mylar & 4 & 3 \\
HDPE (transparent) & 3.0 & 1.6K bolometer, 35\,cm$^{-1}$ & 150Mylar & 4 & 3 \\
Epoxy + stainless steel & 1.0 & 1.6K bolometer, 100\,cm$^{-1}$ & 150Mylar & 4 & 3 \\
Epoxy + stainless steel & 1.0 & 1.6K bolometer, 35\,cm$^{-1}$ & 150Mylar & 4 & 3 \\
Epoxy + stainless steel & 1.5 & 1.6K bolometer, 100\,cm$^{-1}$ & multilayerMylar & 1 & 4 \\
Epoxy + stainless steel & 1.5 & 1.6K bolometer, 35\,cm$^{-1}$ & 125Mylar & 1 & 8 \\
Epoxy + stainless steel & 1.5 & 1.6K bolometer, 35\,cm$^{-1}$ & 50Mylar & 1 & 3 \\
Epoxy + stainless steel & 1.5 & RT-DLaTGS & KBr & 4 & 8 \\
Epoxy + stainless steel & 1.5 & RT-DTGS-FIR & KBr & 4 & 8 \\
Epoxy + stainless steel & 2.0 & 1.6K bolometer, 100\,cm$^{-1}$ & 150Mylar & 4 & 3 \\
Epoxy + stainless steel & 2.0 & 1.6K bolometer, 35\,cm$^{-1}$ & 150Mylar & 4 & 3 \\
Epoxy + stainless steel & 2.0 & RT-DLaTGS & KBr & 4 & 3 \\
Epoxy + stainless steel & 2.0 & RT-DLaTGS & KBr & 4 & 8 \\
Epoxy + stainless steel & 2.0 & RT-DTGS-FIR & KBr & 4 & 3 \\
Epoxy + stainless steel & 2.0 & RT-DTGS-FIR & KBr & 4 & 8 \\
Stycast 2850FT & 1.0 & 1.6K bolometer, 100\,cm$^{-1}$ & 150Mylar & 4 & 3 \\
Stycast 2850FT & 1.0 & 1.6K bolometer, 35\,cm$^{-1}$ & 150Mylar & 4 & 3 \\
Stycast 2850FT & 1.5 & 1.6K bolometer, 100\,cm$^{-1}$ & 150Mylar & 4 & 3 \\
Stycast 2850FT & 1.5 & 1.6K bolometer, 100\,cm$^{-1}$ & multilayerMylar & 1 & 4 \\
Stycast 2850FT & 1.5 & 1.6K bolometer, 35\,cm$^{-1}$ & 125Mylar & 1 & 8 \\
Stycast 2850FT & 1.5 & 1.6K bolometer, 35\,cm$^{-1}$ & 150Mylar & 4 & 3 \\
Stycast 2850FT & 1.5 & 1.6K bolometer, 35\,cm$^{-1}$ & 50Mylar & 1 & 3 \\
Stycast 2850FT & 1.5 & RT-DLaTGS & KBr & 4 & 8 \\
Stycast 2850FT & 1.5 & RT-DLaTGS & KBr & 4 & 8 \\
Stycast 2850FT & 1.5 & RT-DTGS-FIR & KBr & 4 & 8 \\
Stycast 2850FT & 1.5 & RT-DTGS-FIR & KBr & 4 & 8 \\
PTFE & 1.0 & 1.6K bolometer, 100\,cm$^{-1}$ & multilayerMylar & 1 & 4 \\
PTFE & 1.0 & 1.6K bolometer, 35\,cm$^{-1}$ & 125Mylar & 1 & 8 \\
PTFE & 1.0 & 1.6K bolometer, 35\,cm$^{-1}$ & 50Mylar & 1 & 3 \\
PTFE & 1.0 & RT-DLaTGS & KBr & 4 & 8 \\
PTFE & 1.0 & RT-DTGS-FIR & KBr & 4 & 8 \\
PTFE & 1.5 & 1.6K bolometer, 100\,cm$^{-1}$ & multilayerMylar & 1 & 4 \\
PTFE & 1.5 & 1.6K bolometer, 35\,cm$^{-1}$ & 125Mylar & 1 & 8 \\
PTFE & 1.5 & 1.6K bolometer, 35\,cm$^{-1}$ & 50Mylar & 1 & 3 \\
PTFE & 1.5 & RT-DLaTGS & KBr & 4 & 8 \\
PTFE & 1.5 & RT-DTGS-FIR & KBr & 4 & 8 \\
PTFE & 2.0 & 1.6K bolometer, 100\,cm$^{-1}$ & multilayerMylar & 1 & 4 \\
PTFE & 2.0 & 1.6K bolometer, 35\,cm$^{-1}$ & 125Mylar & 1 & 8 \\
PTFE & 2.0 & RT-DLaTGS & KBr & 2 & 8 \\
PTFE & 2.0 & RT-DTGS-FIR & KBr & 2 & 8 \\
UHU plus Endfest 300 & 1.0 & 1.6K bolometer, 100\,cm$^{-1}$ & 150Mylar & 4 & 3 \\
UHU plus Endfest 300 & 1.0 & 1.6K bolometer, 35\,cm$^{-1}$ & 150Mylar & 4 & 3 \\
UHU plus Endfest 300 & 1.0 & RT-DLaTGS & KBr & 4 & 3 \\
UHU plus Endfest 300 & 1.0 & RT-DTGS-FIR & KBr & 4 & 3 \\
UHU plus Endfest 300 & 1.5 & 1.6K bolometer, 100\,cm$^{-1}$ & 150Mylar & 4 & 3 \\
UHU plus Endfest 300 & 1.5 & 1.6K bolometer, 100\,cm$^{-1}$ & multilayerMylar & 1 & 4 \\
UHU plus Endfest 300 & 1.5 & 1.6K bolometer, 35\,cm$^{-1}$ & 125Mylar & 1 & 8 \\
UHU plus Endfest 300 & 1.5 & 1.6K bolometer, 35\,cm$^{-1}$ & 150Mylar & 4 & 3 \\
UHU plus Endfest 300 & 1.5 & 1.6K bolometer, 35\,cm$^{-1}$ & 150Mylar & 4 & 3 \\
UHU plus Endfest 300 & 1.5 & 1.6K bolometer, 35\,cm$^{-1}$ & 50Mylar & 1 & 3 \\
UHU plus Endfest 300 & 1.5 & RT-DLaTGS & KBr & 4 & 8 \\
UHU plus Endfest 300 & 1.5 & RT-DLaTGS & KBr & 4 & 8 \\
UHU plus Endfest 300 & 1.5 & RT-DTGS-FIR & KBr & 4 & 8 \\
UHU plus Endfest 300 & 1.5 & RT-DTGS-FIR & KBr & 4 & 8 \\
UHU plus Endfest 300 & 2.0 & 1.6K bolometer, 100\,cm$^{-1}$ & 150Mylar & 4 & 3 \\
UHU plus Endfest 300 & 2.0 & 1.6K bolometer, 100\,cm$^{-1}$ & multilayerMylar & 1 & 4 \\
UHU plus Endfest 300 & 2.0 & 1.6K bolometer, 35\,cm$^{-1}$ & 125Mylar & 1 & 8 \\
UHU plus Endfest 300 & 2.0 & 1.6K bolometer, 35\,cm$^{-1}$ & 150Mylar & 4 & 3 \\
UHU plus Endfest 300 & 2.0 & RT-DLaTGS & KBr & 4 & 8 \\
UHU plus Endfest 300 & 2.0 & RT-DLaTGS & KBr & 2 & 8 \\
UHU plus Endfest 300 & 2.0 & RT-DTGS-FIR & KBr & 4 & 8 \\
UHU plus Endfest 300 & 2.0 & RT-DTGS-FIR & KBr & 2 & 8 \\
UHU plus Endfest 300 & 2.5 & 1.6K bolometer, 100\,cm$^{-1}$ & multilayerMylar & 1 & 4 \\
UHU plus Endfest 300 & 2.5 & 1.6K bolometer, 35\,cm$^{-1}$ & 125Mylar & 1 & 8 \\
UHU plus Endfest 300 & 2.5 & RT-DLaTGS & KBr & 4 & 8 \\
UHU plus Endfest 300 & 2.5 & RT-DTGS-FIR & KBr & 4 & 8 \\
UHU plus Endfest 300 & 3.0 & 1.6K bolometer, 100\,cm$^{-1}$ & multilayerMylar & 1 & 4 \\
UHU plus Endfest 300 & 3.0 & 1.6K bolometer, 35\,cm$^{-1}$ & 125Mylar & 1 & 8 \\

\hline
\hline
 
\end{longtable}

\end{document}